# Point-contact search for antiferromagnetic giant magnetoresistance

Z. Wei[1], A. Sharma[2], J. Bass[2], and M. Tsoi[1]

[1]University of Texas at Austin, Austin, TX, USA
[2]Michigan State University, East Lansing, MI, USA

## ABSTRACT

We report the first measurements of effects of large current densities on current-perpendicular-to-plane magnetoresistance (MR) of magnetic multilayers containing two antiferromagnetic layers separated by a non-magnetic layer. These measurements were intended to search for a recently predicted antiferromagnetic giant magnetoresistance (AGMR) similar to GMR seen in multilayers containing two ferromagnetic layers separated by a non-magnetic layer. We report on MR measurements for current injected from point contacts into sandwiches containing different combinations of layers of F = CoFe and AFM = FeMn. In addition to: AFM/N/AFM, F/AFM/N/AFM, and F/AFM/N/AFM/F structures, initial results led us to examine also AFM/F/N/AFM, F/AFM, and single F- and AFM-layer structures. At low currents, no MR was observed in any samples, and no MR was observed at any current densities in samples containing only AFMs. Together, these results indicate that no AGMR is present in these samples. In samples containing F-layers, high current densities sometimes produced a small positive MR – largest resistance at high fields. For a given contact resistance, this MR was usually larger for thicker F-layers, and for a given current, it was usually larger for larger contact resistances (smaller contacts). We tentatively attribute this positive MR to suppression at high currents of spin accumulation induced around and within the F-layers.

Giant Magnetoresistance (GMR) in ferromagnetic/nonmagnetic (F/N) multilayers has been a focus of intensive study for two decades, both for interesting fundamental physics [1, 2] and important industrial applications, e.g., read-heads and magnetic memory [3]. In the simplest case, GMR refers to a large change in resistance of an F/N/F trilayer when the relative orientation of the magnetizations of the two F-layers changes from anti-parallel (AP) to parallel (P). Recently a similar effect — antiferromagnetic (AFM) GMR = AGMR — was predicted in structures where F-layers are replaced by AFMs [4]. It was also predicted that a large enough current density passed through an AFM/N/AFM trilayer could change the relative orientation of their magnetizations — AFM spin-transfer-torque (STT) effect — like STT in F/N/F structures [5-7]. Motivated by these predictions, we initiated a search for magnetoresistive and current-induced effects in systems involving two AFM = FeMn = $Fe_{0.5}Mn_{0.5}$ layers separated by a nonmagnetic N=Cu spacer. The predictions were made assuming ballistic transport in perfectly ordered samples. Our experiments are, thus, crucial to see if the effects predicted for idealized 1-dimensional (1D) AFMs can be seen in real samples with diffusive transport and disorder.

In F/N/F trilayers, the relative orientation of the magnetizations of the two Fs is controlled by an externally applied magnetic field B. To achieve well-defined AP and P states, the moment of one of the F-layers is often 'pinned', via exchange coupling (exchange bias) to an adjacent AFM layer [8, 9], leaving the moment of the other free to reverse in much smaller B. In a simple AFM/N/AFM sample, just applying a field B is not expected to be efficient, due to the weak effect of external fields on magnetic moments in AFMs. To achieve better control of the AFMs, we also studied AFM/N/AFM layers sandwiched between two F layers to give F/AFM/N/AFM/F, with the two AFM layers differently exchange coupled to their respective F-neighbors. Applying a magnetic field to change the magnetic order of the F-layers should then also affect the order of the AFM layers. If a large current density is sent through any of these multilayer samples, spin-transfer-torque (STT) interactions between Fs and between AFMs [4, 10] might also affect the AFM magnetic order. Finally to isolate the MR observations of interest from potential spurious effects, we tested a wide variety of structures: AFM/N/AFM, F/AFM/N/AFM, F/AFM/N/AFM/F, AFM/F/N/AFM, F/AFM, and single F- and AFM-layers.

Our multilayers were sputtered onto Si substrates using a system and techniques described in [11]. All samples had a 50 nm thick Cu underlayer to secure a closely perpendicular-to-plane (CPP) current flow, and a 5 nm thick Au capping layer to avoid surface oxidation. The point contacts were made with a standard system [7, 10], using a sharpened Cu wire and a differential screw to move the Cu tip toward the multilayer film. All films used N = 10 nm of Cu, AFM = 3 or 8 nm of FeMn, and F = 2, 3, 4, 6, or 10 nm of CoFe = $Co_{91}Fe_9$ to give different magnitudes of exchange-bias at the AFM/F interfaces and to look for any thickness dependence. All combinations of AFMs (3/N/3, 3/N/8, 8/N/3, 8/N/8) were tested in AFM/N/AFM sandwiches. To induce exchange bias at the F/AFM interfaces, the samples were cooled from ~463K through the Néel temperature of FeMn in a magnetic field ~18 mT. All layer thicknesses are given in nm, and negative current corresponds to electrons flowing from the tip into the multilayer.

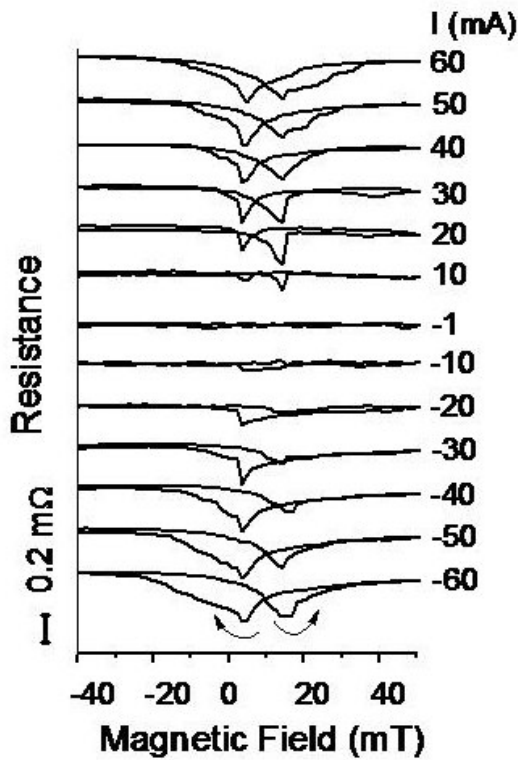

**Fig. 1**. Point-contact magnetoresistance at different bias currents for a 1.3 Ohm point contact to sample CoFe(10)/FeMn(8)/Cu(10)/FeMn(8)/CoFe(3). Solid traces show point-contact resistance R=V/I versus applied magnetic field B for a series of bias currents I = -60 to +60 mA. Arrows indicate up and down B-sweeps. I = 60 mA corresponds to current density j ~ $7 \times 10^{12}$ $A/m^2$. Note that a nonzero magnetoresistance appears only at high bias currents.

For small applied currents, neither standard current-in-plane (CIP) MR measurements on our multilayer films, nor CPP-MR measurements with point contacts, showed any MRs for samples of all types. For larger applied currents, no MR was seen in samples of the form AFM/N/AFM or F/AFM/N/AFM with F-layer thicknesses of only 3 nm. In contrast, Fig. 1 shows sweeps of the point-contact resistance, R=V/I, vs applied field B for a series of currents I applied through a 1.3 Ω contact to a sample of the form F(10)/AFM(8)/N/AFM(8)/F(3). There is no MR for small I. But above a minimum I (both positive and negative), small, positive MRs appear and grow in magnitude with increasing I (sometimes tending toward saturation in magnitude above a certain value of I). For the sweeps shown, starting from high positive field, R(B) is constant at a maximum value, decreases to a minimum at B ~ 5 mT, and then grows again to its maximum value at high negative field. The reversed sweeps from high negative to high positive fields show similar behavior with minimum R(B) at about 15 mT.

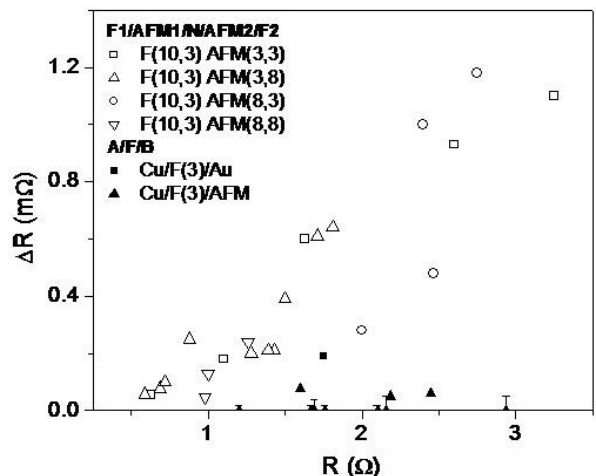

**Fig. 2**. The maximum change in resistance, ΔR, versus point contact resistance, R, at saturation, recorded at I = 30 mA for down-sweeps. Open symbols show data for four different samples of type F1/AFM1/N/AFM2/F2. The legend indicates the thicknesses of the two F (F1, F2) and two AFM (AFM1, AFM2) layers in nanometers. Note that error bars are smaller than symbol sizes. For comparison, filled symbols show data for samples with a single 3 nm thick F layer sandwiched between Cu and Au layers (filled squares) or between Cu and AFM layers (filled triangles).

Similar small spin-valve MR signals (~0.02-0.12%) were seen in 28 contacts (R ranging from 0.6-3.5 Ω) to F1/AFM/N/AFM/F2 type samples with 10 nm thick F1 and 3 nm thick F2. Values of the maximum change in contract resistance ΔR = R(max) – R(min), taken for down sweeps at I = +30 mA, are plotted vs saturation R in Fig. 2 (open symbols) for contacts to all different samples of type F1/AFM/N/AFM/F2 — designated via F1/F2=10/3 and AFM/AFM thicknesses: 8/8, 3/3, 8/3, and 3/8. Overall, ΔR for a given I increases with the increasing contact resistance (decreasing contact size), and similar values of ΔR are seen for samples with different AFM thicknesses. In contrast, only one of six contacts to sample of type F1/AFM/N/AFM/F2 with both F1 and F2 =3 nm thick showed a positive MR comparable with that found in samples with at least one F(10 nm) layer. The other five contacts all produced data points below the distribution shown in Fig. 2 with open symbols. This observation suggests that the observed MR might be associated with single F(10 nm) layers in our samples. To check this possibility, we measured an extensive number of samples that include only one F layer each (either 3 or 10 nm thick) sandwiched between Au, Cu, or AFM layers, sometimes with an extra AFM layer: AFM/F/N/AFM, F/AFM/N/AFM, N/F/AFM, and N/F/N. Figure 2 (filled symbols) shows that, with one exception, contacts to samples with only one F=3 nm layer gave MRs much smaller than those for the open symbols. In contrast, Fig. 3 shows that most contacts to samples with one F=10 nm layer (filled symbols) gave positive MRs comparable in magnitude to those of the open symbols.

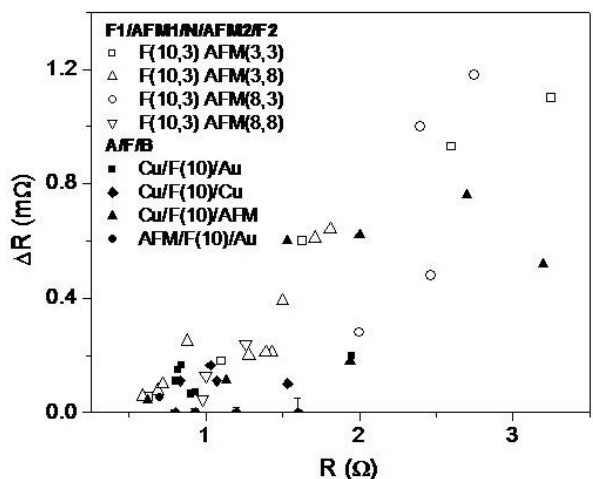

**Fig. 3**. The maximum change in resistance, ΔR, versus point contact resistance, R, at saturation, recorded at I = 30 mA for down-sweeps. Open symbols show the same data for F1/AFM1/N/AFM2/F2 samples as in Fig. 2. Filled symbols show data for samples with a single 10 nm thick F layer sandwiched between Cu and Au (filled squares), Cu and Cu (filled diamonds), Cu and AFM (filled triangles), AFM and Au (filled circles). Note that error bars are smaller than symbol sizes.

To summarize: (i) we observe small positive MRs (resistance is highest at saturation) in samples of type F/AFM/N/AFM/F when at least one of the Fs is 10 nm thick; (ii) the MR is present only at high current densities flowing across such multilayers (no MR at small currents); (iii) only one out of six contacts to a sample with two 3 nm thick F-layers showed comparable MR; (iv) samples of types AFM/F/N/AFM, F/AFM/N/AFM, N/F/AFM, or N/F/N with only one F=3 nm show no MR (with exception of only one contact), but similar samples with F=10 nm show (in most cases) MRs comparable to those in (i).

The positive MR in Fig. 1 cannot be standard GMR between the two outer F-layers, which must be negative (smallest R at large B). The absence of MR in F(3)/AFM/N/AFM samples also rules out anisotropic MR, which is independent of F-layer thickness. The observation of MR in F/AFM/N/AFM/F samples might be tentatively attributed to the AGMR predicted in [4]. In support, are a correlation of MR-shape with SQUID-measured magnetizations of our samples (not shown) and the need for high currents to change the AFM order parameters [4, 10]. However, the presence of similar MRs in samples with only a single F(10 nm) layer (no AFMs) suggests that the MR is more likely associated with Fs in our multilayers. Small positive MRs were previously seen in nanopillars of single F=Permalloy layers [12]. There they were associated with suppression at high currents of spin accumulation induced around and within the F-layer. At high Bs, the magnetization of F is uniform and the pillar resistance is higher due to an extra contribution from spin accumulation [12,13]. Near zero B, however, the Oersted field of the applied current produces a vortex in F on a scale comparable to the spin diffusion length. This non-uniform magnetization suppresses the spin accumulation, decreasing the pillar resistance. A similar mechanism may be responsible for the positive MRs of our multilayers. If so, this would be the first evidence for such behavior in extended layers.

The authors thank S. Urazhdin for suggesting that the point contact to extended-layer data shown in Figs. 1 & 2 might involve similar behavior to that previously seen in nanopillars [12]. This work was supported in part by NSF grants DMR-06-45377 and DMR-08- 04126.

[1] M. N. Baibich et al., Phys. Rev. Lett. **61**, 2472 (1988).
[2] G. Binasch, P. Grünberg, F. Saurenbach, W. Zinn, Phys. Rev. B **39**, 4828 (1989).
[3] J. M. Daughton, IEEE Trans. Magn. **36**, 2773 (2000).
[4] A. S. Núñez, R. A. Duine, P. Haney, A. H. MacDonald, Phys. Rev. B **73**, 214426 (2006).
[5] J. C. Slonczewski, J. Magn. Magn. Mater. **159**, L1 (1996).
[6] L. Berger, Phys. Rev. B **54**, 9353 (1996).
[7] M. Tsoi et al., Phys. Rev. Lett. **80**, 4281 (1998).
[8] W. H. Meiklejohn, C. P. Bean, Phys. Rev. **102**, 1413 (1956).
[9] J. Stöhr, H. C. Siegmann, *"Magnetism: from fundamentals to nanoscale dynamics"* (Springer Series in Solid-State Sciences, Vol. 152, Springer, 2006).
[10] Z. Wei et al., Phys. Rev. Lett. **98**, 116603 (2007).
[11] J. M. Slaughter, W. P., Pratt, Jr, P. A. Schroeder, Rev. Sci. Instrum. **60**, 127 (1989).
[12] S. Urazhdin, C. L. Chien, K. Y. Guslienko, L. Novozhilova, Phys. Rev. B **73**, 054416 (2006).
[13] P. C. van Son, H. van Kempen, and P. Wyder, Phys. Rev. Lett. **58**, 2271 (1987).